\documentclass[preprint,10pt]{sigplanconf}

\usepackage{wrapfig}
\usepackage{amsmath}
\usepackage{amssymb}
\usepackage{array}
\usepackage{url}
\usepackage{stmaryrd}
\usepackage{listings}

\usepackage{lmodern}
\usepackage[T1]{fontenc}

\usepackage{graphicx}
\usepackage{color}
\usepackage{times}

\usepackage{listings}
 
\lstdefinelanguage{scala}{
  morekeywords={%
          script,then,here,there,% added for SubScript
          abstract,case,catch,class,def,do,else,extends,%
          false,final,finally,for,forSome,if,implicit,import,lazy,%
          match,new,null,object,override,package,private,protected,%
          return,sealed,super,this,throw,trait,true,try,type,%
          val,var,while,with,yield},
  otherkeywords={<<, >>, ==>, =>, ->, +\~\~, +\~/\~>, +\~/\~, \~\~, \~/\~>, \~/\~, \~>,<\%,<:,>:,\#,@,<=,<-, =, \{!, !\}, \{, \}},
  sensitive=true,
  morecomment=[l]{//},
  morecomment=[n]{/*}{*/},
  morestring=[b]",
  morestring=[b]',
  morestring=[b]"""
}[keywords,comments,strings]
 
\newcommand{\code}[1]{{\texttt{#1}}}

\newcommand*{\last}{\hspace{-3pt}\leftarrow\hspace{-7pt}*}
\newcommand*{\rast}{*\hspace{-7pt}\rightarrow\hspace{-3pt}}
\newcommand*{\lrast}{\hspace{-3pt}\leftarrow\hspace{-7pt}*\hspace{-7pt}\rightarrow\hspace{-3pt}}

\newcommand*{\parallelO}{\hspace{-1.6pt}\bigcirc\hspace{-10.0pt}\parallel\hspace{2.8pt}}

\newcommand*{\astOt}{\hspace{-1.6pt}\bigcirc\hspace{-7.0pt}*\hspace{2.8pt}}

\newcommand*{\acpinterrupt}{\mapsto\hspace{-3.4pt}\vartriangleright}

\begin{document}

\CopyrightYear{2015}

\title{Some New Directions for ACP Research}

\authorinfo{Andr\'{e} van Delft}
  {Independent Researcher}
  {andre dot vandelft at gmail dot com}
\maketitle

%%%%%%%%%%%%%%%%%%%%%%%%%%%%%%%%%%%%%%%%%%%%%%%%%%%%%%%
\begin{abstract}
%%%%%%%%%%%%%%%%%%%%%%%%%%%%%%%%%%%%%%%%%%%%%%%%%%%%%%%
This paper lists some new directions for research related to the Algebra of Communicating Processes (ACP).
Most of these directions have  been inspired by work on SubScript, an ACP based extension to the programming language Scala.
SubScript applies several new ideas that build on ACP, but currently these lack formal treatment.
\\ 
Some of these new ideas are rather fundamental. 
E.g. it appears that the theory of ACP may well apply to structures of any kind of items, rather than to just processes.
\\
The aim of this list is to raise awareness of the research community about these new ideas; this could help both the research area and the programming language SubScript.
\end{abstract}

%\category {D.1.6}{Logic Programming}
%\category {D.3.1}{Programming Languages}{Formal Definitions and Theory}
%\category {D.3.3}{Language Constructs and Features}{Concurrent programming structures}
 


%%%%%%%%%%%%%%%%%%%%%%%%%%%%%%%%%%%%%%%%%%%%%%%%%%%%%%%
%\vspace{-6pt}
\section{Introduction}
%%%%%%%%%%%%%%%%%%%%%%%%%%%%%%%%%%%%%%%%%%%%%%%%%%%%%%%

The Algebra of Communicating Processes (ACP) is a concurrency theory that allows for concise specifications of event-driven and concurrent processes. It also helps formal reasoning about process behavior. ACP is good at describing processes that communicate synchronously, and less at describing asynchronously communicating processes.
\\
The theory has been developed with mathematical rigor, mainly in the period 1982 to the end of the 20th century. In spite of its robustness ACP is not as much being used in the software engineering world as it could be. This is pitiful since both worlds could profit from exchanges of ideas.
\\
ACP is well applicable as a basis to extend existing programming with nondeterministic expressiveness. We are developing an ACP based extension to the programming language Scala by the name of SubScript
\footnote{Web site: www.subscript-lang.org}.
In a way SubScript walks ahead of ACP: it applies several ideas that would likely fit in ACP, but are not yet formalized.
\\
We think such formalizations are due for a formal specification of SubScript's semantics. These could also bring new life into ACP research. 
\\
The rest of this section introduces ACP and SubScript\footnote{
These sections contain some text fragments literally copied or adapted from a paper presented at the Scala Workshop 2013 \cite{vanDelft:2013:DCL:2489837.2489849} about dataflow programming support in SubScript.};
In the next section we present the list of new directions. For each element we give an impression of the formal treatment that we foresee.

%%%%%%%%%%%%%%%%%%%%%%%%%%%%%%%%%%%%%%%%%%%%%%%%%%%%%%%
%\vspace{-6pt}
\subsection{ACP}
%%%%%%%%%%%%%%%%%%%%%%%%%%%%%%%%%%%%%%%%%%%%%%%%%%%%%%%

The Algebra of Communicating Processes \cite{Baeten:2005:ACP} is an algebraic approach to reasoning about concurrent systems. It is a member of the family of mathematical theories of concurrency known as process algebras or process calculi\footnote{This description of ACP has largely been taken from Wikipedia.}.
More so than the other seminal process calculi (CCS \cite{Milner:1982:CCS} and CSP \cite{Hoare:1985:CSP}), the development of ACP focused on the algebra of processes, and sought to create an abstract, generalized axiomatic system for processes.
\newline
ACP uses instantaneous, atomic actions (a,b,c,...) as its main primitives. Two special primitives are the deadlock process $\delta$ and the empty process $\epsilon$. Expressions of primitives and operators represent processes. The main operators can be roughly categorized as providing a basic process algebra, concurrency, and communication:

\begin{itemize}
\item \emph{Choice and sequencing} - the most fundamental of algebraic operators are the alternative operator ($+$), which provides a choice between actions, and the sequencing operator ($\cdot$), which specifies an ordering on actions. So, for example, the process $(a+b)\cdot{}c$ first chooses to perform either a or b, and then performs action c. How the choice between a and b is made does not matter and is left unspecified. Note that alternative composition is commutative but sequential composition is not (because time flows forward).
\item \emph{Concurrency} - to allow the description of concurrency, ACP provides the merge operator $\parallel$. This represents the parallel composition of two processes, the individual actions of which are interleaved. As an example, the process $(a \cdot b)\parallel(c\cdot d)$ may perform the actions a, b, c, d in any of the sequences abcd, acbd, acdb, cabd, cadb, cdab.
\item \emph{Communication} - pairs of atomic actions may be defined as communicating actions, implying they can not be performed on their own, but only together, when active in two parallel processes. This way, the two processes synchronize, and they may exchange data.
\end{itemize}

ACP fundamentally adopts an axiomatic, algebraic approach to the formal definition of its various operators. Using the alternative and sequential composition operators, ACP defines a basic process algebra which satisfies the following axioms:
\begin{eqnarray*}
x+y &=& y+x\\
(x+y)+z &=& x+(y+z)\\
x+x &=& x\\
(x+y)\cdot z &=& x \cdot z + y \cdot z\\
(x \cdot y) \cdot z &=& x \cdot (y \cdot z)  
\end{eqnarray*}

The primitives $0$ and $1$, also known as $\delta$ and $\epsilon$, behave much like the 0 and 1 that are usually neutral elements for addition and multiplication in algebra:
\begin{eqnarray*}
0+x &=& x          \\
0 \cdot x &=& \delta          \\
1 \cdot x &=& x          \\
x \cdot 1 &=& x          
\end{eqnarray*}
There is no axiom for $x \cdot 0$. \\

$x+1$ means: \textit{optionally x}. This is illustrated by rewriting $(x+1) \cdot y$ using the given axioms:
\begin{eqnarray*}
(x+1) \cdot y  &=&  x \cdot y +1 \cdot y  \\
&=&  x \cdot y + y
\end{eqnarray*}

The parallel merge operator $\parallel$ is defined in terms of the alternative and sequential composition operators. This definition also requires two auxiliary 
operators:
\begin{eqnarray*}
x\parallel y  =  x\llfloor y + y\llfloor x + x|y
\end{eqnarray*}

\begin{itemize}
\item $x\llfloor y$ - "left-merge": $x$ starts with an action, and then the rest of x is done in parallel with $y$. 

\item $x|y$ - "communication merge": $x$ and $y$ start with a communication (as a pair of atomic actions), and then the rest of $x$ is done in parallel with the rest of $y$.
\end{itemize}

The definitions of many new operators such as the left merge operator use a special property of closed process expressions with $\cdot$ and $+$: with the axioms as term rewrite rules from left to right (except for the commutativity axiom for $+$), each such expression reduces into one of the following normal forms: $(x + y)$, $a \cdot x$, $1$, $0$. E.g. the axioms for the left merge operator are:
\begin{eqnarray*}
(x+y) \llfloor z &=& x \llfloor z + y \llfloor z \\
a \cdot x \llfloor y &=& a \cdot (x \llfloor y) \\
1 \llfloor x &=& 0          \\
0 \llfloor x &=& 0          
\end{eqnarray*}
Again these axioms may be applied as term rewrite rules so that each closed expression with the parallel merge operator $\parallel$ reduces to one of the four normal forms. This way it has been possible to extend ACP with many new operators that are defined precisely in terms of sequence and choice, e.g. interrupt and disrupt operators, process launching, and notions of time and priorities.
\\
Since its inception in 1982, ACP has successfully been applied to the specification and verification of among others, communication protocols, traffic systems and manufacturing plants. 
\newline
In 1989, Henk Goeman unified Lambda Calculus with process expressions \cite{DBLP:journals/ipl/Goeman90}. Shortly thereafter, Robin Milner et al developed Pi-calculus \cite{Milner89acalculus}, which also combines the two theories.

%%%%%%%%%%%%%%%%%%%%%%%%%%%%%%%%%%%%%%%%%%%%%%%%%%%%%%%
%\vspace{-6pt}
\subsection{SubScript}
%%%%%%%%%%%%%%%%%%%%%%%%%%%%%%%%%%%%%%%%%%%%%%%%%%%%%%%

SubScript extends the Scala language with refinement constructs that are called "scripts". Essentially these scripts are definitions of ACP-like processes.
Fragments of Scala code that are enclosed in brace pairs serve as atomic actions. These fragments serve as operands in script expressions, together with script calls, iterator operands and others. The symbol for normal parallel composition is \code{\&}; in practice this is not as much used as "or-parallelism" denoted by \code{||}.
\\
More on SubScript is explained using following examples are typical for SubScript; these apply most of the new ideas that this paper is about.

%%%%%%%%%%%%%%%%%%%%%%%%%%%%%%%%%%%%%%%%%%%%%%%%%%%%%%%
%\vspace{-6pt}
\subsection{Example: A GUI Application}
%%%%%%%%%%%%%%%%%%%%%%%%%%%%%%%%%%%%%%%%%%%%%%%%%%%%%%%

Suppose we need a simple program to look up items in a database, based on a search string.

\begin{figure}[hb]
\includegraphics[scale=0.60]{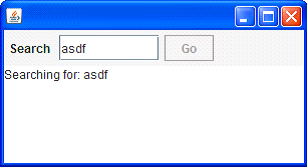}
\end{figure}

The user can enter a search string in the text field and then press the Go button. This will at first put a "SearchingÉ" message in the text area at the lower part. Then the actual search will be performed at a database, which may take a few seconds (simulated by a call to Thread.sleep). Finally the results from the database are shown in the text area.

In SubScript this could look like\footnote{The SubScript website shows versions of the same functionality in plain Java and Scala, that are much less concise and intuitive).}:
\\
\begin{lstlisting}
 live              = searchSequence ...

 searchSequence    = searchCommand 
                     showSearchingText 
                     searchInDatabase 
                     showSearchResults
                     
searchCommand = clicked(searchButton)
\end{lstlisting}

\begin{itemize}

\item The white space after \code{searchSequence} in the \code{live} script denotes sequential composition. Semicolons also mean sequential composition\footnote{These have a low operator priority.}. For new lines there are similar rules as in Scala, implying that these often denote sequential composition as well\footnote{In SubScript these new lines have an even lower operator priority.}.

\item The three dots in the \code{live} script (\code{...}, "etcetera") turn the main script into an "eternal" sequential loop of search sequences.

\item \code{searchCommand} represents the command that the user gives to make the search start. It comes down to clicking a button. 

\item \code{showSearchingText} and \code{showSearchResults} are scripts that sets texts in the larger text area.\\\code{searchInDatabase} performs a search in the background; meanwhile the GUI remains active. The exact definitions of these 3 scripts are beyond the scope of this paper.

\end{itemize}

Scala has an "implicit conversion" mechanism, which can apply a method call to a value that appears in a program text, when such a method is in scope been marked as \code{implicit}. SubScript extends this mechanism by including script calls as well. So if the script \code{clicked} has been marked as \code{implicit}, its name may be left out, leaving:

\begin{lstlisting}
searchCommand = searchButton
\end{lstlisting}

%%implicit
%%parallel or >> style
%%iterations
%%disruption and interruption
%%data flow
%%lambda
%comm, disamb., negation

\subsection{Extending the program}
Now we add some realistic requirements to the program.
\begin{itemize}
\item The search action may also be triggered by the user pressing the Enter key in the search text field.
\item The search action requires that the input text field is not empty; only then should the search button be enabled
\item The user should be able to cancel an ongoing search, by clicking a Cancel button, or pressing the Escape key. 
\item The user can exit the application by clicking an Exit button, or by clicking in the close box at the window's upper right corner. But exiting should first be confirmed in a dialog box. 
\end{itemize}
\begin{figure}[hb]
\includegraphics[scale=0.60]{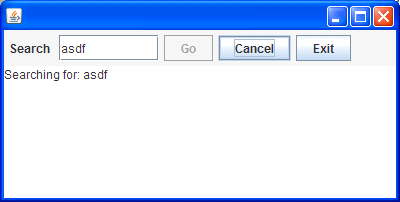}
\end{figure}

The 3 user commands will be:
\begin{lstlisting}
searchCommand = searchButton + Key.Enter
cancelCommand = cancelButton + Key.Escape
exitCommand   = exitButton   + windowClosing
\end{lstlisting}
The first and second plus operators create exclusive choices between buttons and key codes. These operands are not processes, but data items for which implicit conversions to processes have been defined (such as \code{clicked} and \code{keyPressed}).
\\
Script \code{windowClosing} acts on window closing events.
\\
Exiting is implemented using a process named \code{exit} that runs in or-parallel composition to the rest. The or-parallel operator is \code{||}, it means that all operands happen; as soon as one finishes successfully then the other is terminated and the whole composition terminates successfully.
In this case, the left hand operand is an eternal loop of search sequences; the right hand operand is a (probably) finite loop. \\
\\
The \code{exit} process starts with the exit command being given; then a confirmation dialog is run; all to be repeated while the result of the confirmation dialog is false. The result of the confirmation dialog is transferred using a dataflow operator to a while construct; this operator is a curly arrow that names and types the flowing data item.
\begin{lstlisting}
live = searchSequence... || exit

exit = exitCommand 
       confirmExit ~~(b:Boolean)~~> while(!b)
\end{lstlisting}

(\code{!} is the logical negation operator in Scala)
\\
For the search sequence we now add items at the start and the end.
\\\\
\code{searchGuard} is  an "active guard" containing a sequential loop. 
It first checks whether the text field contains some text. If so, there is an "optional break", specified by the dot. This means that the sequence and thus also the guard may end successfully, so that \code{searchCommand} becomes active. \\
However maybe an event happens at the text field before the user issues this search command; then the check needs to be redone, etc (\code{...}). 
\\
After \code{searchCommand} a new line follows; this separates the first line from the remaining five lines. Therefore the rest, including \code{cancelSearch}, can only become active after the \code{searchCommand} has happened. \code{cancelSearch} is preceded by a slash symbol, which stands for disruption: the left hand side happens, possibly disrupted when the right hand side starts happening. The parentheses group the items on the preceding lines, so that the whole becomes the left hand side of the slash operator.
\begin{lstlisting}
searchSequence = searchGuard searchCommand 
                 ( showSearchingText 
                   searchInDatabase 
                   showSearchResults
                  )
                 / cancelSearch
                 
searchGuard     = if(!searchTF.text.isEmpty) .
                  anyEvent(searchTF) 
                  ...
                   
cancelSearch = cancelCommand showCanceledText
\end{lstlisting}

%%%%%%%%%%%%%%%%%%%%%%%%%%%%%%%%%%%%%%%%%%%%%%%%%%%%%%%
%\vspace{-6pt}
\section{New Directions}
%%%%%%%%%%%%%%%%%%%%%%%%%%%%%%%%%%%%%%%%%%%%%%%%%%%%%%%

%%%%%%%%%%%%%%%%%%%%%%%%%%%%%%%%%%%%%%%%%%%%%%%%%%%%%%%
\subsection{Algebra of Items}
%%%%%%%%%%%%%%%%%%%%%%%%%%%%%%%%%%%%%%%%%%%%%%%%%%%%%%%

In the previous example, buttons, key codes and event codes were added:

\begin{lstlisting}
searchCommand = searchButton + Key.Enter
cancelCommand = cancelButton + Key.Escape
exitCommand   = exitButton   + windowClosing
\end{lstlisting}
 
Thanks to Scala's \code{implicit} mechanism we can glue any kind of items together using the ACP operators. 
The result is an algebra of general items rather than just of processes.
\\
A transformation may turn such an item specification to a process specification. In the above, the obvious transformation are about receiving input. It is also possible to do the opposite: the specification may for instance instruct a test robot to generate button clicks etc.
\\
A particular application of such algebra's of items is in language grammars: these are descriptions of certain text structures, rather than processes that accept or produce such texts.
\\
The paper "Equivalence notions for concurrent systems and refinement of actions"\cite{vanGlabbeekGoltz:1989} already covers some technical issues of such a generalization to an algebra of items.

%%%%%%%%%%%%%%%%%%%%%%%%%%%%%%%%%%%%%%%%%%%%%%%%%%%%%%%
\subsection{Generalized Communication}
%%%%%%%%%%%%%%%%%%%%%%%%%%%%%%%%%%%%%%%%%%%%%%%%%%%%%%%

The Algebra of Communicating Processes (ACP) \cite{Baeten:2005:ACP} offers process communication through atomic actions that combine into other atomic actions. We may call this action-level communication. The binary action-level communication generalizes to synchronous communication with more than two parties. Namely, if $a$ and $b$ are such actions, and if $a|b = d$ and also $d|c = e$, then obviously $(a|b)|c = e$.
\\
In practice though communication between processes may last longer than just a single atomic action. In general multiple processes will share the execution of a sequence of actions, according to another process specification\footnote{This has been implemented in a predecessor language of SubScript; for SubScript communication is still due at the time of writing.}. Like in ACP with action-level communication, a generalization to more than 2 communicating parties was possible in that language.
\\
Now when communication of 2 or more parties is possible, why not say it is about 1 or more parties? The 1-party case then coincides with the notion of process refinement. 
\\
In traditional programming languages refinements are known as functions, procedures and methods. For these there is always 1 caller that calls a callee. In a process oriented language more than one caller may be required to call a callee, in a synchronous effort. For instance, a process describing a message transmission may have to be called by two parties: a sender and a receiver.
\\
Conceptually this generalization of normal refinements to synchronous communication with 1 or more parties is simple and elegant. A formal treatment of this general communication seems to be well possible, though considerably more complicated than action-level communication. 

%%%%%%%%%%%%%%%%%%%%%%%%%%%%%%%%%%%%%%%%%%%%%%%%%%%%%%%
\subsection{Combination with Lambda Calculus}
%%%%%%%%%%%%%%%%%%%%%%%%%%%%%%%%%%%%%%%%%%%%%%%%%%%%%%%

The presented \code{exit} script contains a dataflow arrow:
\begin{lstlisting}
exit = exitCommand 
       confirmExit ~~(b:Boolean)~~> while(!b)
\end{lstlisting}

The left-hand side of the arrow operator needs to produce a boolean value for transmission to the right hand side. 
Technically this requires, among others, support for "anonymous scripts" 
or, to use another phrase, "anonymous processes" or "process lambda's". 
These are much comparable to "lambda's" in functional programming languages, 
also known as anonymous functions.
\\
In the \code{exit} script the call to \code{confirmExit} is technically 
an anonymous script, and the same holds for \code{while(!b)}.
\\
As stated in the section on ACP, both Milner and Goeman combined lambda calculus with process calculus.
However, this was done for the process calculus CCS. 
The theoretic treatment of a combination of lambda calculus with ACP is still lacking.

%%%%%%%%%%%%%%%%%%%%%%%%%%%%%%%%%%%%%%%%%%%%%%%%%%%%%%%
\subsection{Result values, exceptions and dataflow}
%%%%%%%%%%%%%%%%%%%%%%%%%%%%%%%%%%%%%%%%%%%%%%%%%%%%%%%

There is more to the previous dataflow example. In the subexpression
\begin{lstlisting}
confirmExit ~~(b:Boolean)~~> while(!b)
\end{lstlisting}

the left hand side, \code{confirmExit}, is to produce a boolean value 
for transport via the arrow to the right hand side.
This requires a new notion: it should be possible to give a process may have a result value, 
which becomes available when the process has success.
\\
But such a process may also fail, e.g. when something goes wrong. 
In terms of modern programming languages that would coincide with an exception being thrown.
Hence the result of a process may not only be the prospective yield, but also an exception.
\\
It is possible to specify how both the normal result value and the exception should flow.
E.g.,
\begin{lstlisting}
 x ~~(b:Boolean)~~> y
 +~/~(e:Exception)~~> z
\end{lstlisting}

This means in SubScript: do script \code{x}; when this succeeds continue with \code{y}, handing it the yielded value. 
In case \code{x} ends in failure, probably an exception has resulted; then continue with \code{z} with the exception.
\\
Under the hood this uses a ternary construct
\begin{lstlisting}
do x then y else z
\end{lstlisting}
A counterpart for this has not yet been specified for ACP; this does appear to be relatively straightforward.
\\
A formal treatment of result values for processes in ACP and their flows, will probably require much more effort.

%%%%%%%%%%%%%%%%%%%%%%%%%%%%%%%%%%%%%%%%%%%%%%%%%%%%%%%
\subsection{Iterations and their termination}
%%%%%%%%%%%%%%%%%%%%%%%%%%%%%%%%%%%%%%%%%%%%%%%%%%%%%%%

There are multiple forms of iterations for ACP processes.
\\
For a statically known "iterative sum" the symbol $\sum$ is used with annotated limits, as in common mathematical texts.
Likewise for iterative sequences there is $\prod$, and occasionally for parallel iterations a large version of $||$ exists.
\\
A more dynamic approach is the use of recursive specifications, as in $X = a\cdot{X} + b$, which denotes a sequence of zero or more $a$'s, terminated by a $b$. 
\\
Lastly there are variations of the Kleene star in use, e.g. $a*b$, meaning the same as the foregoing recursive specification. Here the Kleene star is an operator, just like $+$ and $ \cdot{} $.
\\\\
SubScript has recursion, but not the other two options. Instead to support many kinds of iterations, there are various special operands, that may be used in conjunction with any operator such as \code{+}, \code{;} and \code{||}. Some of these operands have already been introduced in the examples:
\\
\begin{tabular}[b]{cp{6.7cm}}\vspace{5pt}
\code{while}   & marks a loop and a conditional mandatory break \\\vspace{5pt}
 \code{for}      & much like \code{while} and like the Scala for-comprehension \\\vspace{5pt}
 \code{...}       & marks a loop; no break point, at least not here \\\vspace{5pt}
 \code{..}        & marks a loop, and at the same time an optional break \\\vspace{5pt}
 \code{.}         & an optional break point \\\vspace{5pt}
 \code{break} & a mandatory break point \\
\end{tabular}
\\
These operands would need some formal treatment. Note that some are in fact easily defined in terms of others:
\begin{itemize}
\item\code{..} is a combination of \code{...} and \code{.}
\item\code{while(b)} is a combination of \code{...} and \\\code{if (!b) break}.
\end{itemize}

The meaning of the ellipsis operand (\code{...}) seems to be relatively easy to define. 
This is by defining a transformation of an expression with such an operand into an expression without such an operand.
\\
Consider for any operator, say $*$ (which can actually be \code{+}, \code{;} etc), a process $x_1 * x_2 * ... * x_n$.
(Here these tiny dots stand for $x_i$ terms).
If any of these terms is the ellipsis operand, then
\begin{itemize}
\item the entire expression should be replaced by the solution of the equation $X = x_1 * x_2 * ... * x_n * X$
\item in this equation all occurrences of  the ellipsis operand should be replaced by a either 0 or 1, whichever is the neutral operand for $*$. 
\end{itemize} 

The treatment of \code{break} and \code{.} is harder. This is mainly due to the fact that these have informally been given meanings that depend on the governing operator. The differentiation appeared to be necessary to maximize applicability.
\\

%%%%%%%%%%%%%%%%%%%%%%%%%%%%%%%%%%%%%%%%%%%%%%%%%%%%%%%
\subsection{Operator axioms style}
%%%%%%%%%%%%%%%%%%%%%%%%%%%%%%%%%%%%%%%%%%%%%%%%%%%%%%%

ACP axiomatizations of parallelism usually starts with an axiom applying two auxiliary operators:
\begin{eqnarray*}
x\parallel y  =  x\llfloor y + y\llfloor x + x|y
\end{eqnarray*}
Here $x\llfloor y$ means, informally speaking: the left hand side $x$ should start with an atomic action; then the rest of $x$ happens in parallel with the right hand side $y$.
\\
And $x|y$ means: first $x$ and $y$ start to communicate by executing a shared atomic action; thereafter the a parallel composition of the rest of $x$ and $y$ happens.

A parallel composition may terminate successfully if both operands may terminate successfully. Therefore from the axiomatization the following should be deductible:
\begin{eqnarray*}
1\parallel 1  =  1
\end{eqnarray*}

There is a problem here: should the $1$ be produced in rules for the $x\llfloor y$ part or for the $x|y$ part? Both do not seem to fit: the former form is meant to express that one party starts with an atomic action; the latter is meant to express that both parties start by communicating with one another.
\\
This inconvenience is reflected in the treatment of the $1$ in ACP publications: some let the $1$ be produced in the $x\llfloor y$ part, others apply the $x|y$ part.
\\
A better approach may well be to introduce another auxiliary operator specifically meant to for producing $1$'s, as in
\begin{eqnarray*}
x\parallel y  &=&  x{\parallelO}y + x\llfloor y + y\llfloor x + x|y
\end{eqnarray*}

This way a minor change in the definition of $x{\parallelO}y$ could yield a kind of or-parallelism: successful termination may occur when any of the two operands may succeed successfully.
\\
Another fruitful change would be the introduction of an auxiliary right-merge operator.

\begin{eqnarray*}
x\parallel y  &=&  x{\parallelO}y + x\llfloor y + x\rrfloor y + x|y
\end{eqnarray*}

This is really not needed for the parallel operator; but the right-merge operator is easily defined in terms of the left-merge operator.

\begin{eqnarray*}
x\rrfloor y &=& y\llfloor x
\end{eqnarray*}

The main use for this style with an extra auxiliary operator is to enable a uniform approach for more operators, that may or may not be symmetric. In general defining axioms would start with something like
\begin{eqnarray*}
x * y  &=&  x{\astOt}y + x\ {\last}\ y + x\ {\rast}\ y + x\lrast y
\end{eqnarray*}

Here the asterisk is stands for any operator to be defined this way; the asterisk with arrows denote auxiliary versions for left-composition, right-composition and communication.

%%%%%%%%%%%%%%%%%%%%%%%%%%%%%%%%%%%%%%%%%%%%%%%%%%%%%%%
\subsection{And- and or-parallelism}
%%%%%%%%%%%%%%%%%%%%%%%%%%%%%%%%%%%%%%%%%%%%%%%%%%%%%%%

In SubScript programs the or-parallel operator \code{||} occurs much more than the normal parallel operator \code{\&}.
The total family consists of 4 operators, two with and-like logic behavior and two with or-like logic behavior:
\\
\code{\&}, \code{\&\&}, \code{|}, \code{||}
\\
\\
The reason to have four parallel operators may come clear by an analogy with the same symbols for boolean expression operators, as used in programming languages C, C++, C\#, Java and Scala:
\\
In those languages, the evaluation of boolean expressions \code{x\&y} and \code{x|y} requires that both operands are evaluated.
\\
For the evaluation of boolean expressions \code{x\&\&y} and \code{x||y} first the left hand side operand is evaluated; only if the result of that evaluation would not be decisive for the logic result, the right hand side is evaluated. So if for instance \code{x} evaluates to \code{true} in \code{x\&\&y} then \code{y} is evaluated, and that result sets the result of the whole.
\\
Likewise the process operators \code{\&\&} and \code{||} have disruptive capabilities:
\begin{itemize}
\item\code{||} terminates one operand as soon as the other finishes successfully, and then the whole succeeds.
\item Likewise the and-parallel operator \code{\&\&} terminates one operand as soon as the other terminates with failure.
\end{itemize}

The axiomatic definition of these operators is helped by the style introduced in the previous subsection. The definition with structural operational semantics (SOS) requires negative premises. It seems to be rather complicated to prove that the axiomatic and SOS definitions conform to one another. It is not yet clear whether current automatic proof systems are capable to handle this.

%%%%%%%%%%%%%%%%%%%%%%%%%%%%%%%%%%%%%%%%%%%%%%%%%%%%%%%
\subsection{Interruption and disruption}
%%%%%%%%%%%%%%%%%%%%%%%%%%%%%%%%%%%%%%%%%%%%%%%%%%%%%%%

ACP literature defines "Mode transfer" operators to express disruption and interruption \cite{Baeten00modetransfer}. Unfortunately these definitions do not handle the cases that on operator equals $1$. Worse, the provided definition for the interrupt operator can not be extended to handle the $1$.\footnote{The underlying cause is that ACP researchers often studied minimal systems that did not contain $1$. This was because smaller systems may be studied with more rigor. Also some ACP researchers had a low relatively appreciation for the $1$, not realizing that it is able to express optionality in sequences, as shown in subsection 1.1.}.
\\
Namely, when $1$ is added to the algebra, $a\cdot{1} = a$.
Therefore axioms LINT1 and LINT2 in the paper by Baeten and Bergstra would imply $1\acpinterrupt{a} = 1$. 
However, from INT+RINT: $1 \acpinterrupt a = something + a$, which contradicts the previous equality.
\\
A related problem of the current ACP definition of interruption is that it is optional, e.g, $a\acpinterrupt{b} = a + b\cdot{a}$.
This is not natural. If an interrupt is optional, why should after such an interrupt no other interrupts be allowed any more?
Moreover there is no way of expressing that an interrupt is mandatory.
\\
Therefore SubScript will have a mandatory interrupt operator. To make it optional, just add 1 to the right hand side: 

\begin{lstlisting}
 x %/ (y+1)
\end{lstlisting}

But since it is natural to express that a process may be interrupted 0 or more times, sequentially, SubScript has also an operator for that\footnote{The notation may look strange; these operators belong to a family of "suspend\&resume" operators; their symbols all start with \code{\%}. The suspend/resume operators have not yet been implemented in SubScript.}:
\begin{lstlisting}
 x %/% y
\end{lstlisting}
 
This "multi interruptions" operator also needs a formal definition. 
\\
A particularity is that it does not have a neutral element. That is unfortunate, since various constructs in SubScript should behave more or less neutrally, in specific respects. For instance, an \code{if-then} construct that has no explicit \code{else} clause behaves as if it has an implicit {else} clause that behaves neutrally. Also the iteration operand \code{...} behaves neutrally, apart from the fact that it turns its governing operator into an iteration.
\\
Although no neutral element exists for \code{\%/\%}, it seems reasonably safe to make constructs such as \code{...} behave as $0$ when they are governed by the multi-interruptions operator, apart from their specific aspects.

%%%%%%%%%%%%%%%%%%%%%%%%%%%%%%%%%%%%%%%%%%%%%%%%%%%%%%%
\subsection{Stream flow}
%%%%%%%%%%%%%%%%%%%%%%%%%%%%%%%%%%%%%%%%%%%%%%%%%%%%%%%

A special kind of dataflow may be defined for listening on a stream.
Suppose \code{s} is a process that listens to a stream; every time a datum arrives from the stream an atomic action in \code{t} happens. Now at such occasions a script \code{p} should be executed on those data. This could be expressed as an iterating data flow operator:
\begin{lstlisting}
s ~~(d:Datum)~~>> t
\end{lstlisting}
The definition would require a "mandatory interruptions" operator\footnote{Like other suspend/resume operators this has not yet been implemented in SubScript.}:
\begin{lstlisting}
 x %/%/ y
\end{lstlisting}
This performs \code{x} with the modification that each time that \code{x} succeeds \code{y} happens. When \code{y} succeeds then \code{x} may resume with its next action, etc; if there is no such action the entire process succeeds.
\footnote{An earlier version of this paper stated a different meaning: "This performs \code{x} with the modification that each time after an atomic action in \code{x} happens, \code{y} happens". However, for the use in a dataflow operator, it is necessary that the left hand side produces a value that flows to the right; this only happens when the left hand side succeeds}
\\
A stream pipeline streams would be possible as in: 
\begin{lstlisting}
s ~~(d:Datum)~~>> t ~~(d:Datum)~~>> u
\end{lstlisting}
To make sense, the left hand side of the second arrow should not include \code{s}, since otherwise \code{u} would happen too many times. Therefore this flow operator should be right associative.
\\
To merge to streams \code{s} and \code{t} with this flow operator would be like 
\begin{lstlisting}
(s&t) ~~(d:Datum)~~>> u
\end{lstlisting}
Splitting a stream would be like 
\begin{lstlisting}
s ~~(d:Datum)~~>>( t ~~(d:Datum)~~>> u
                 & v ~~(d:Datum)~~>> w)
\end{lstlisting}
It seems that such stream features might both be relatively easy to add on ACP, after other forms of dataflow have been added; they may also lead to clear and appealing specifications.

%%%%%%%%%%%%%%%%%%%%%%%%%%%%%%%%%%%%%%%%%%%%%%%%%%%%%%%
\subsection{Negation}
%%%%%%%%%%%%%%%%%%%%%%%%%%%%%%%%%%%%%%%%%%%%%%%%%%%%%%%

ACP may be viewed as an extension to Boolean Algebra. 
The values F (False) and T (True) would become 0 and 1 in ACP.
The atomic actions are then a new class of basic element; they have a notion of time
or at least sequence, and this breaks the commutativity of the multiplication.
\\
However, next to addition and multiplication, Boolean Algebra has also a negation operator.
This suggests that ACP could also have such an operator. A simple definition would be that 
the negation of a process performs the actions of the process; only:
\\when the original process may terminate successfully, the negation does not
\\when the original process terminates without success, the negation terminates successfully.
\\
This is easily defined using the ternary \code{do}-\code{then}-\code{else} construct, that was touched upon in the Dataflow subsection:
\begin{lstlisting}
 -x = do x then 0 else 1
\end{lstlisting}

In a different definition be like the previous one, with the following extra rule:
\\the negated process may terminate successfully whenever the original process negation cannot do so.
\\
Only rarely has such a construct been needed as a programming feature\footnote{These unary operators had been implemented in SubScript's predecessor language, but not yet in SubScript itself.}.

%%%%%%%%%%%%%%%%%%%%%%%%%%%%%%%%%%%%%%%%%%%%%%%%%%%%%%%
\subsection{Disambiguated choice}
%%%%%%%%%%%%%%%%%%%%%%%%%%%%%%%%%%%%%%%%%%%%%%%%%%%%%%%

Ambiguous choices are an issue in parsing texts according to given grammars.
The following 3 grammars will lead to ambiguous choices:
\begin{lstlisting}
  A B A C + A D
 (A B + 1)  A D
  A B / A C
\end{lstlisting}
Unambiguous versions would respectively look like:
\begin{lstlisting}
  A (B A C + D)
  A (B A D + D)
  A (B + C + A C)
\end{lstlisting}

Here the common summand \code{A} has been factored out. Such a transformation is most often possible,
but some grammars look more natural with ambiguous choices. 
\\
A first option to evade the ambiguous choice in SubScript would be to specify an or-parallel operator instead of the nondeterministic choice: \code{|} or \code{||}. For instance, in
\begin{lstlisting}
  A B A C || A D
\end{lstlisting}
the \code{A} could be programmed in such a way that multiple instances can succeed given the corresponding input. Next, when \code{B} would succeed, \code{D} should fail and vice versa. This way, either the \code{A B A C} part or the \code{A D} part would succeed. The other or-parallel operator \code{|} could be used as well. But there is a third option. A disambiguating choice operator \code{|+|} could be defined so that the equivalence
\begin{lstlisting}
  A B A C |+| A D  =  A (B A C + D)
\end{lstlisting}
would hold.
\\
An ACP paper by Baeten and Mauw \cite{Baeten:Mauw:1995} presents an approach for such an operator. It is defined using three auxiliary operators.
\\
This approach seems unfit for implementation in SubScript; the main problem would be to determine when two atomic actions are essentially the same. A more natural way to specify such relationships is using communicating scripts. In particular, it will be possible in SubScript to define a script of which one or more instances may communicate with one another\footnote{Communication has not yet been implemented in SubScript.}:
\begin{lstlisting}
 script s,.. = scriptBody
\end{lstlisting}

Now the disambiguating choice operator \code{|+|} is much like the exclusive-or operator \code{+}, the difference being that an atomic action in one operand need not exclude an atomic action in another operand, in case they communicate or have another kind of simultaneous occurrence. 
\\
But this would not be all. There would also be a need for a disambiguating version of the sequence operator. Namely
\begin{lstlisting}
  (A B |+| 1)  A D
\end{lstlisting}
would simply reduce to
\begin{lstlisting}
  (A B + 1)  A D
\end{lstlisting}

We would need operators \code{|;}, \code{||;}, \code{|;|} that would give the following reductions:
\begin{lstlisting}
  (A B + 1)|;  A D  =  A (B A D |   D)
  (A B + 1)||; A D  =  A (B A D ||  D)
  (A B + 1)|;| A D  =  A (B A D |+| D)
\end{lstlisting}
Similarly a disambiguating version \code{|/|} of the disrupt operator \code{/} would be needed.
\\
\\
For each of these disambiguating operators, implementation and formal definition seem to be well feasible.

%%%%%%%%%%%%%%%%%%%%%%%%%%%%%%%%%%%%%%%%%%%%%%%%%%%%%%%
%\vspace{-6pt}
\section{Conclusion}
%%%%%%%%%%%%%%%%%%%%%%%%%%%%%%%%%%%%%%%%%%%%%%%%%%%%%%%

We have presented a set of new ideas for ACP research.
This list is not final; the work on SubScript will likely cause more such items to emerge.
\\
Adopting these ideas would make ACP wider applicable, while at the same time it would
give a solid foundation for their counterparts in SubScript.
\\
ACP researchers who wish to coauthor publications in any of these directions are invited to contact the author.

%%%%%%%%%%%%%%%%%%%%%%%%%%%%%%%%%%%%%%%%%%%%%%%%%%%%%%%
%\vspace{-6pt}
\bibliographystyle{abbrvnat}
% The bibliography should be embedded for final submission.
%\nocite{*}
\bibliography{bibdb}

\begin{thebibliography}{9}
\providecommand{\natexlab}[1]{#1}
\providecommand{\url}[1]{\texttt{#1}}
\expandafter\ifx\csname urlstyle\endcsname\relax
  \providecommand{\doi}[1]{doi: #1}\else
  \providecommand{\doi}{doi: \begingroup \urlstyle{rm}\Url}\fi

\bibitem[Baeten and Bergstra(2000)]{Baeten00modetransfer}
J.~Baeten and J.~Bergstra.
\newblock Mode transfer in process algebra.
\newblock Technical report, 2000.

\bibitem[Baeten and Mauw(1995)]{Baeten:Mauw:1995}
J.~Baeten and S.~Mauw.
\newblock Delayed choice: an operator for joining massage sequence charts.
\newblock In \emph{Formal Description Techniques VII (Proceedings of the 7th
  IFIP WG 6.1 International Conference, Berne, Switzerland, 1994)}, pages
  340--354, 1995.

\bibitem[Baeten(2005)]{Baeten:2005:ACP}
J.~C.~M. Baeten.
\newblock A brief history of process algebra.
\newblock \emph{Theor. Comput. Sci.}, 335:\penalty0 131--146, May 2005.

\bibitem[Goeman(1990)]{DBLP:journals/ipl/Goeman90}
H.~Goeman.
\newblock Towards a theory of (self) applicative communicating processes: A
  short note.
\newblock \emph{Inf. Process. Lett.}, 34\penalty0 (3):\penalty0 139--142, 1990.

\bibitem[Hoare(1985)]{Hoare:1985:CSP}
C.~Hoare.
\newblock Communicating sequential processes.
\newblock \emph{ACM Computing Surveys}, 7\penalty0 (1):\penalty0 80--112, 1985.

\bibitem[Milner(1982)]{Milner:1982:CCS}
R.~Milner.
\newblock \emph{A Calculus of Communicating Systems}.
\newblock Springer-Verlag New York, Inc., Secaucus, NJ, USA, 1982.

\bibitem[Milner et~al.(1989)Milner, Parrow, and Walker]{Milner89acalculus}
R.~Milner, J.~Parrow, and D.~Walker.
\newblock A calculus of mobile processes, part i.
\newblock \emph{I AND II. INFORMATION AND COMPUTATION}, 100, 1989.

\bibitem[van Delft(2013)]{vanDelft:2013:DCL:2489837.2489849}
A.~van Delft.
\newblock Dataflow constructs for a language extension based on the algebra of
  communicating processes.
\newblock In \emph{Proceedings of the 4th Workshop on Scala}, SCALA '13. ACM,
  2013.

\bibitem[van Glabbeek and Goltz(1989)]{vanGlabbeekGoltz:1989}
R.~van Glabbeek and U.~Goltz.
\newblock Equivalence notions for concurrent systems and refinement of actions.
\newblock In A.~Kreczmar and G.~Mirkowska, editors, \emph{Mathematical
  Foundations of Computer Science 1989}, volume 379 of \emph{Lecture Notes in
  Computer Science}, pages 237--248. Springer Berlin Heidelberg, 1989.
\newblock \doi{10.1007/3-540-51486-4_71}.

\end{thebibliography}
%%%%%%%%%%%%%%%%%%%%%%%%%%%%%%%%%%%%%%%%%%%%%%%%%%%%%%%

%%%%%%%%%%%%%%%%%%%%%%%%%%%%%%%%%%%%%%%%%%%%%%%%%%%%%%%
%\begin{table*} - for single column; unfortunately this would break the 10-page limit

\vspace{66pt}

\end{document}